\documentclass[epjST]{svjour}
\usepackage{cite}
\usepackage{graphics}
\usepackage[dvips]{graphicx}
\usepackage{dcolumn}
\usepackage{bm}
\usepackage{dsfont}
\usepackage{color}
\usepackage[normalem]{ulem} 
\usepackage{url}
\usepackage[colorlinks=true ,citecolor=blue]{hyperref}
\usepackage{amsmath,amssymb}

\newcommand{\be}{\begin{equation}}
\newcommand{\ee}{\end{equation}}
\newcommand{\bea}{\begin{eqnarray}}
\newcommand{\eea}{\end{eqnarray}}
\newcommand{\mv}[1]{\langle #1\rangle}
\newcommand{\f}{\frac}

\newcommand{\ket}{\rangle}
\newcommand{\ra}{\rightarrow}

\begin{document}
\title{Two-body bound and edge states in the extended SSH Bose-Hubbard model}
\author{M. Di Liberto\inst{1}\fnmsep\thanks{\email{marco.diliberto@unitn.it}} \and
  A. Recati\inst{1,2} \and I. Carusotto\inst{1} \and C. Menotti\inst{1} }
\institute{INO-CNR BEC Center and Dipartimento di Fisica, Universit\`a di Trento, 38123 Povo, Italy \and 
Arnold Sommerfeld Center for Theoretical Physics, Ludwig-Maximilians-Universit\"at M\"unchen, 80333 M\"unchen,
  Germany}
\abstract{ We study the bosonic two-body problem in a
  Su-Schrieffer-Heeger dimerized chain with on-site and
  nearest-neighbor interactions. We find two classes of bound
  states. The first, similar to the one induced by on-site
  interactions, has its center of mass on the strong link, whereas the
  second, existing only thanks to nearest-neighbors interactions, is
  centered on the weak link. We identify energy crossings between
  these states and analyse them using exact diagonalization and
  perturbation theory. In the presence of open boundary conditions,
  novel strongly-localized edge-bound states appear in the spectrum as
  a consequence of the interplay between lattice geometry, on-site and
  nearest-neighbor interactions. Contrary to the case of purely
  on-site interactions, such EBS persist even in the strongly
  interacting regime.} %end of abstract
\maketitle

\section{Introduction}
\label{intro}

A peculiar feature of interacting lattice models is the existence of
two-particle bound states for both attractive and repulsive
interactions. This can be intuitively understood as a consequence of
the bounded kinetic energy bandwidth, which does not allow dissipation
of large interaction energies
\cite{MattisRMP,winkler2006,DaleyRev,Valiente2009,Valiente2010,Valiente2011,minganti2016}. For
repulsive interactions, such doublons are therefore stable high-energy
composite objects with their own dynamics and properties. In the
presence of nearest-neighbor interactions, the picture becomes even
richer because stable objects can be formed by two particles sitting
on neighboring sites \cite{Valiente2009,Valiente2010,alba2013,Marques2016}.

In recent years, the study of topological states of matter has
disclosed a novel class of fascinating quantum phases, now well
understood and classified at the single-particle level
\cite{KaneRMP10}. A key feature of these states is the presence of
edge modes that are robust and protected by symmetries. Much less
understood is the role of interactions on these phases. On one hand,
interactions can be responsible of very non-trivial many-body
topological states, as the notorious fractional Quantum Hall effect
\cite{Laughlin1983}. On the other hand, interactions can have a
detrimental effect on the single-particle topological phases
\cite{Schnyder2016}.

In Ref.~\cite{PRA}, we have shown that non-trivial physics arises
for two interacting particles in the simplest topological lattice model,
namely the Su-Schrieffer-Heeger chain, despite the fact that two-body
on-site interactions reduce the symmetries protecting the
single-particle edge modes. Indeed, for moderate values of on-site
interactions, two-body edge modes may be found, the origin of which is
attributed to the interplay between lattice geometry, topology and
interactions themselves (see also \cite{Gorlach2016}). These states
can be understood as interaction-induced Tamm-Schockley surface modes
\cite{Tamm32,Shockley39}.

In this work, we extend our previous analysis including also
nearest-neighbor interactions. We identify a new class of two-body
out-of-cell bound states and study the resonances appearing in the
spectrum when the energy of the new dimers matches the energy of the
in-cell bound states studied in \cite{PRA}. For finite chains, the
presence of two-body edge bound states (EBS) is revealed in both
dimerizations, with novel and enhanced localization properties due to
nearest-neighbor interactions. We analyze the properties of bound and
edge-bound states using exact diagonalization and effective theories
in the strong-dimerization or strong-interaction limit.

Upon a mapping of the two-body wave function in  one
  dimension (1D) onto a free particle wave function in an
appropriately engineered two-dimensional (2D) lattice
\cite{Longhi2011,Corrielli2013,PRA}, on-site interactions in 1D are
translated onto an energy off-set in the main diagonal of the 2D
lattice. Nearest-neighbor interactions can be analogously translated
onto energy off-sets in the adjacent diagonals. In optical fiber
setups, this mapping can be implemented through refractive-index
modulations.  At the two-body level, this makes it possible to realize
nearest-neighbor interactions of arbitrary intensity, much harder to
obtain in experiments with ultracold atoms \cite{Ferlaino2015}.
Hence, optical fiber setups are probably the best candidates to
explore experimentally our findings
\cite{Schreiber2012,Mukherjee2015,Mukherjee2016}.

The paper is organized as follows. In Sec.~\ref{sec:model}, we
introduce our model describing interacting particles in a dimerized
lattice of alternating strong and weak links. In Sec.~\ref{sec:dimer},
we describe two classes of dimer states appearing when the strong
links are decoupled from each other. We relate the first class to the
{\it in-cell} dimer states found in \cite{PRA} and discuss the
completely different nature of the second class of {\it out-of-cell}
dimers induced by nearest-neighbor interactions. In
Sec.~\ref{sec:bound}, we introduce a finite weak tunneling to allow
dimer mobility and create delocalized two-body bound states. In
particular, we discuss the resonances occurring between the two
classes of bound states. In Sec.~\ref{sec:edge}, we focus on the case
of open boundary conditions and analyze the effect of
nearest-neighbor interactions on the two-body  bound
edge states. In Sec.~\ref{sec:conclusions}, we draw our conclusions.

%%%%%%%%%%%%%%%%%%%%%%%%%
\section{Model Hamiltonian}
\label{sec:model}

\begin{figure}
\center
\includegraphics[width=.7\columnwidth]{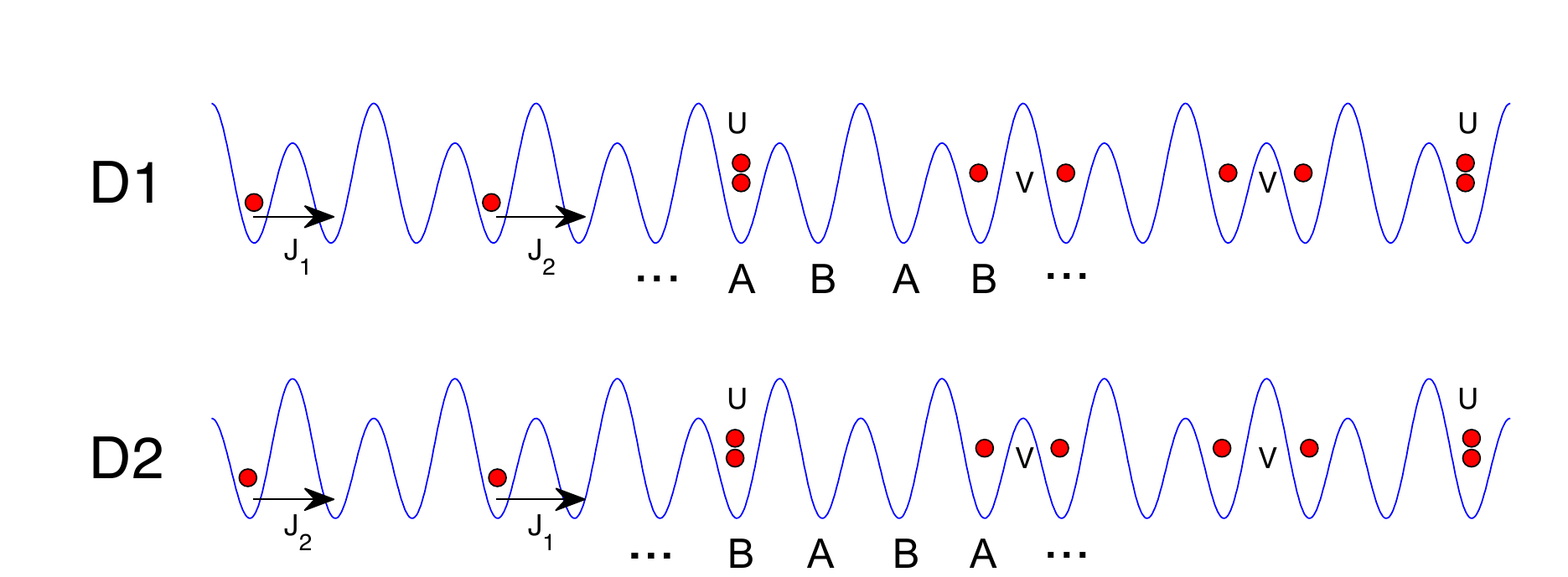}
\caption{Sketch of the SSH model with on-site and nearest-neighbor
  interactions. For open boundary conditions (OBC) and even number of
  lattice cells, one can have two type of dimerizations D1 and D2
  depending on whether the chain ends with strong or weak links.}
\label{fig:0}      
\end{figure}

The Su-Shrieffer-Heeger model (SSH) describes a single particle moving
on a one-dimensional lattice where the nearest-neighbor hopping
coefficients alternate in magnitude. The single-particle SSH Hamiltonian reads \cite{Su1979}

\begin{equation}
\label{SSH}
H^{}_0 = - J_1 \sum_i c^\dag_{A,i} c^{}_{B,i} - J_2 \sum_i
c^\dag_{A,i+1} c^{}_{B,i} + \rm{H.c.}  \,.
\end{equation}
We consider bosonic particles interacting with on-site interactions
\begin{equation}
\label{int_hamU}
H^{}_U = \f{U}{2} \sum_i  \left[ n_{A,i}(n_{A,i}-1) + n_{B,i}(n_{B,i}-1)  \right]
\end{equation}
and nearest-neighbor interactions
\begin{equation}
\label{int_hamV}
H^{}_V = V \sum_i \left( n_{A,i}n_{B,i} + 
  n_{A,i+1}n_{B,i} \right)\,,
\end{equation}
where $n_{\sigma,i} \equiv c^\dag_{\sigma,i}c^{}_{\sigma,i}$ is the
density operator, $\sigma=A,B$ indicates the lattice site and $i$
labels the lattice cell (see Fig.~\ref{fig:0}).

In this work, we focus on the two-body physics in the
strongly-dimerized regime, namely $J_2 \ll J_1$. This naturally allows
us to define as {\it strong link} those pairs of lattice sites $A_i$
and $B_i$ belonging to the same lattice cell $i$, and as {\it weak
  link} those pairs of neighboring lattice sites $B_i$ and $A_{i+1}$
belonging to two neighboring lattice cells.  The interplay between
in-cell kinetic energy $H_{J_1} = - J_1 \sum_i c^\dag_{A,i} c^{}_{B,i}
+ \rm{H.c.}$ and interactions determines the two-body states in the
fully dimerized case ($J_2=0$). The effect of the weak tunneling
Hamiltonian between different cells $H_{J_2}=- J_2 \sum_i
c^\dag_{A,i+1} c^{}_{B,i} + \rm{H.c.}$ will be considered exactly in
the numerical simulations and introduced analytically at the
perturbative level.  For open boundary conditions (OBC) and even
number of cells, two dimerizations are possible, as sketched in
Fig.~\ref{fig:0}(D1,D2).

%
%%%%%%%%%%%%%%%%%%%%%%%%%%%
\section{Dimer states} 
\label{sec:dimer}

First, we consider the fully dimerized case of $J_2 = 0$. Bound pairs
can be formed in the same lattice cell, as in the case of on-site
interactions \cite{PRA, ValienteEPL}, or in neighboring lattice cells,
exclusively induced by nearest-neighbor interactions.

As previously introduced in \cite{PRA}, the Hilbert space of the
strong-link {\it in-cell} Hamiltonian is spanned by the states
$|A_iA_i\ket$, $|A_i B_i\ket$ and $|B_iB_i\ket$. In the presence of
on-site and nearest-neighbors interactions, the Hamiltonian reads
\be
H^{\textrm{cell}}_i =
\begin{pmatrix}
U & -\sqrt{2} J_1 & 0 \\
-\sqrt{2} J_1 & V & -\sqrt{2} J_1\\
0 & -\sqrt{2} J_1 & U
\end{pmatrix}\,.
\label{H_cell}
\ee
This Hamiltonian determines three dimer states $d_\alpha$,
respectively at energies $\epsilon_{1,3} = \f 1 2 \left( U + V \mp
\sqrt{16J_1^2 +(U-V)^2}\right)$ and $\epsilon_2 = U$. The nature of
these states is similar to the one discussed in \cite{PRA} for
$V=0$. In fact, as far as the in-cell dimer wave functions are
concerned, $V$ induces nothing else than a renormalized on-site
interaction $U-V$. Hence, a repulsive nearest-neighbor interaction $V$
favours on-site occupation in $d_1$ and off-site occupation in $d_3$,
leaving $d_2$ unaltered.

Dimer states of completely different nature are induced by the
presence of nearest-neighbor interaction between two atoms in
neighboring cells.  The Hilbert space properly taking into account
{\it out-of-cell} dimers is spanned by the states $[|A_iA_{i+1}\ket +
  |B_iB_{i+1}\ket]/\sqrt{2}$, $|A_iB_{i+1}\ket$ and $|B_iA_{i+1}\ket$.
We do not include in this subspace the fourth out-of-cell state
$[|A_iA_{i+1}\ket - |B_iB_{i+1}\ket]/\sqrt{2}$ at zero energy, which
is decoupled by the rest.  In the considered subspace, the Hamiltonian
reads
\be
H^{\textrm{nn-cells}}_{i+1/2} =
\begin{pmatrix}
0 & -\sqrt{2} J_1 & -\sqrt{2} J_1  \\
-\sqrt{2} J_1 & 0 & 0 \\
-\sqrt{2} J_1 & 0 & V 
\end{pmatrix}\,,
\label{H_nncells}
\ee
where the subscript $i+1/2$ indicates the center-of-mass of the
out-of-cell dimers.
The out-of-cell Hamiltonian predicts three dimer states $|\tilde
d_\beta\ket$\footnote{Out-of-cell dimer states are reminiscent of the
  out-of-cell $d_{NN}$ state discussed in \cite{PRA}, which was formed
  by an effective nearest-neighbor interaction due to second-order
  coupling between different dimer states (see \cite{PRA}, Appendix
  3).  In the presence of a small but finite nearest-neighbor
  interaction $V$, state $d_{NN}$ is replaced by
  $\tilde{d}_3$.}. Their energies $\tilde\epsilon_\beta$ cannot be
computed analytically, but two important limits can be identified: (i)
for $V \to 0$, one finds $\tilde\epsilon_\beta=[-2 J_1, 0, 2J_1]$ and
the corresponding eigenstates tend to states of the three (type I)
scattering continua around energy $\tilde\epsilon_\beta$; (ii) for $V
\to \infty$, one finds $\tilde\epsilon_\beta=[-\sqrt{2} J_1, \sqrt{2}
  J_1, V]$ corresponding to the eigenstates $[1;-1;0]$, $[1;1;0]$, and
$[0;0;1]$ written in our basis. The latter, $|B_i,A_{i+1}\ket$, is
obviously the state which maximizes nearest-neighbor interaction
energy and is the off-site nearest-neighbor analog of the two in-cell
states $d_2$ and $d_3$ in the large $U$ limit. The two states at $\pm
\sqrt{2} J_1$ are instead the out-of-cell analog of state $d_1$ at
$V=0$, minimizing the interaction energy. The effect of
nearest-neighbor interactions on such states of two delocalized
particles in neighboring cells is to enhance the occupation of one
particle in the outer most site of one cell, leaving the second
particle delocalized in the other cell. Due to the suppression of
nearest-neighbor occupation, the kinetic energy is reduced to the
asymptotic value of $\tilde\epsilon_{1,2}=\pm\sqrt{2}J_1$.

It is useful to notice that at $U=0$, the six dimer states always
appear in the sequence $[\tilde\epsilon_1, \epsilon_1, \epsilon_2,
  \tilde\epsilon_2, \tilde\epsilon_3, \epsilon_3]$. Knowing the
asymptotic values of the energies as a function of $U$ and $V$, one
can predict the presence of energy level crossings.

%%%%%%%%%%%%%%%%%%%%%
\section{Bound states}
\label{sec:bound}

\begin{figure}
\center
\includegraphics[width=.65\columnwidth]{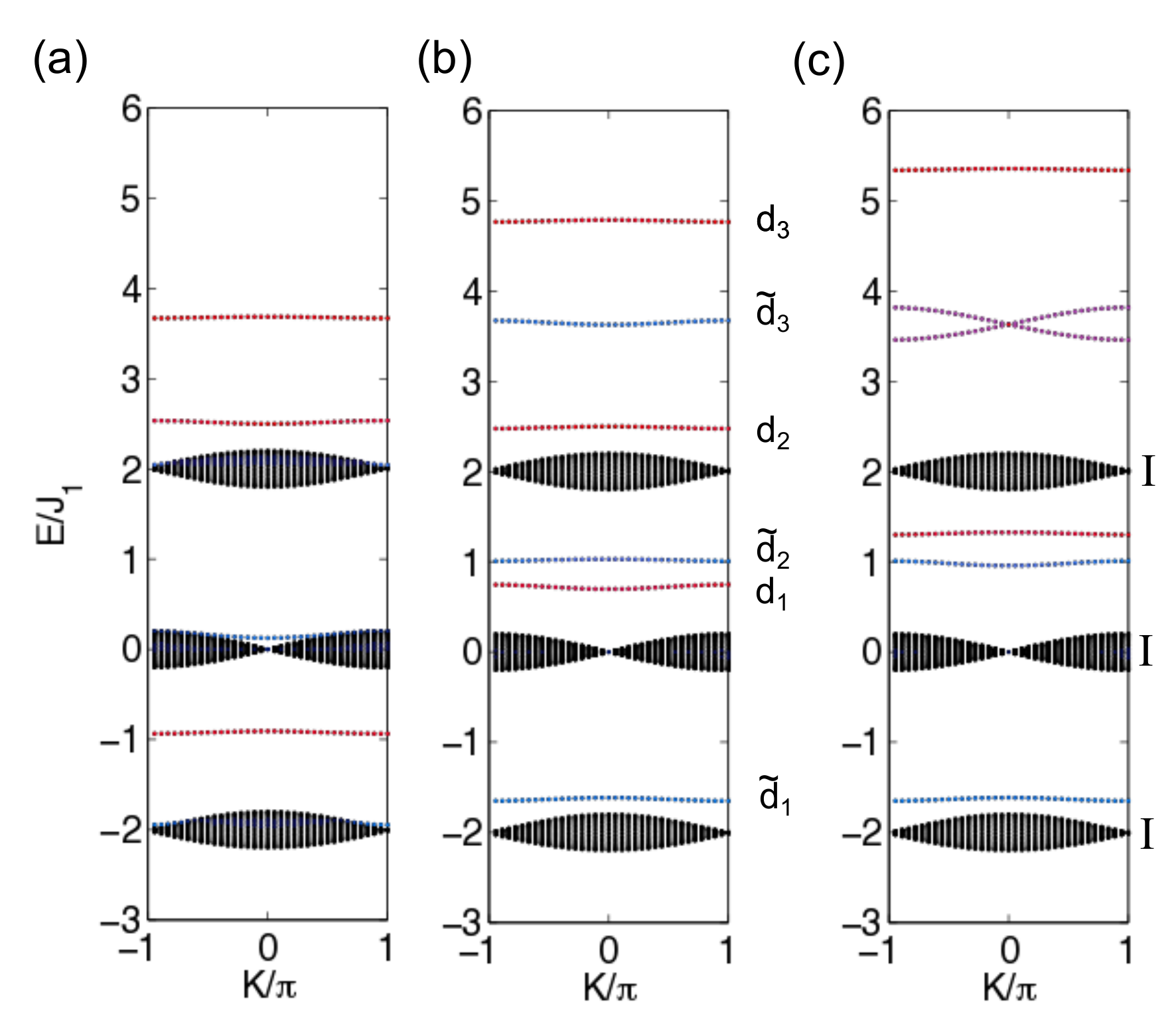}
\caption{Two-body energy spectrum for PBC as a function of the center
  of mass momentum $K$ for 72 lattice sites (36 lattice cells) for
  $J_2=0.1 J_1$, $U=3.63 J_1$ and different values of $V$: (a) $V=0.3
  J_1$; (b) $V=2J_1$; (c) $V=3J_1$. Red colorscale indicates relative
  in-cell population, while blue colorscale indicates relative
  nearest-neighbor out-of-cell population.}
\label{fig:3}    
\end{figure}

\begin{figure}
\center
\includegraphics[width=.65\columnwidth]{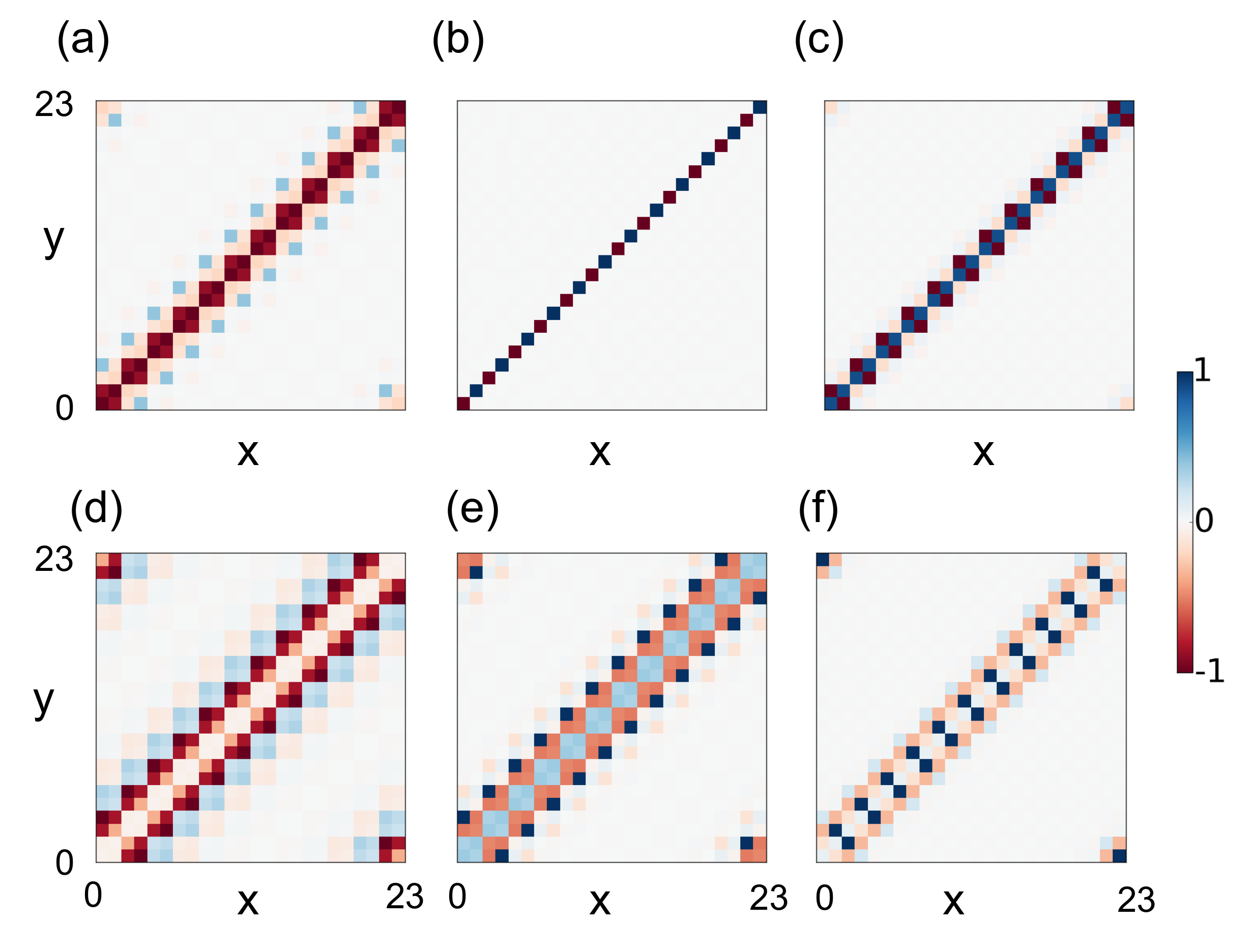}
\caption{(a-c) Bound state wavefunction $|d_1\ket$, $|d_2\ket$,
  $|d_3\ket$ at $K=0$ for $J_1=0.1 J_1$, $U=2.5 J_1$, and $V=3J_1$;
  (d-f) Bound state wavefunctions $|\tilde d_1\ket$, $|\tilde
  d_2\ket$, $|\tilde d_3\ket$ at $K=0$ for $J_2=0.1 J_2$, $U=2.5 J_1$
  and $V=3J_1$. The color code is normalized to the maximum absolute
  value of the wave function.}
\label{fig:2}    
\end{figure}

In the presence of a small but finite hopping $J_2$, the dimers are
allowed to move in the lattice developing narrow bound-state bands.
Consequently, the full two-body energy spectrum - obtained e.g. in
Fig.~\ref{fig:3} by exact diagonalization with periodic boundary
conditions (PBC) - shows three (type I) scattering continua, typical
of the non-interacting two-body SSH model, and six bound-state bands.
For dominant $U$ and small $V$, the out-of-cell bound states may be
well defined only for certain values of the center-of-mass momentum
$K$ either in the Brillouin zone center or boundary (see
Fig.~\ref{fig:3}(a)).  The in-cell and out-of-cell character of the
different bound states is evident in their wave function plotted in
Fig.~\ref{fig:2}.  The bound states $d_\alpha$ and $\tilde d_\beta$
for $U=2.5$ and $V=3$ are respectively shown in Figs.~\ref{fig:2}(a-c)
and Figs.~\ref{fig:2}(d-f).

\begin{figure}
\center
\includegraphics[width=.75\columnwidth]{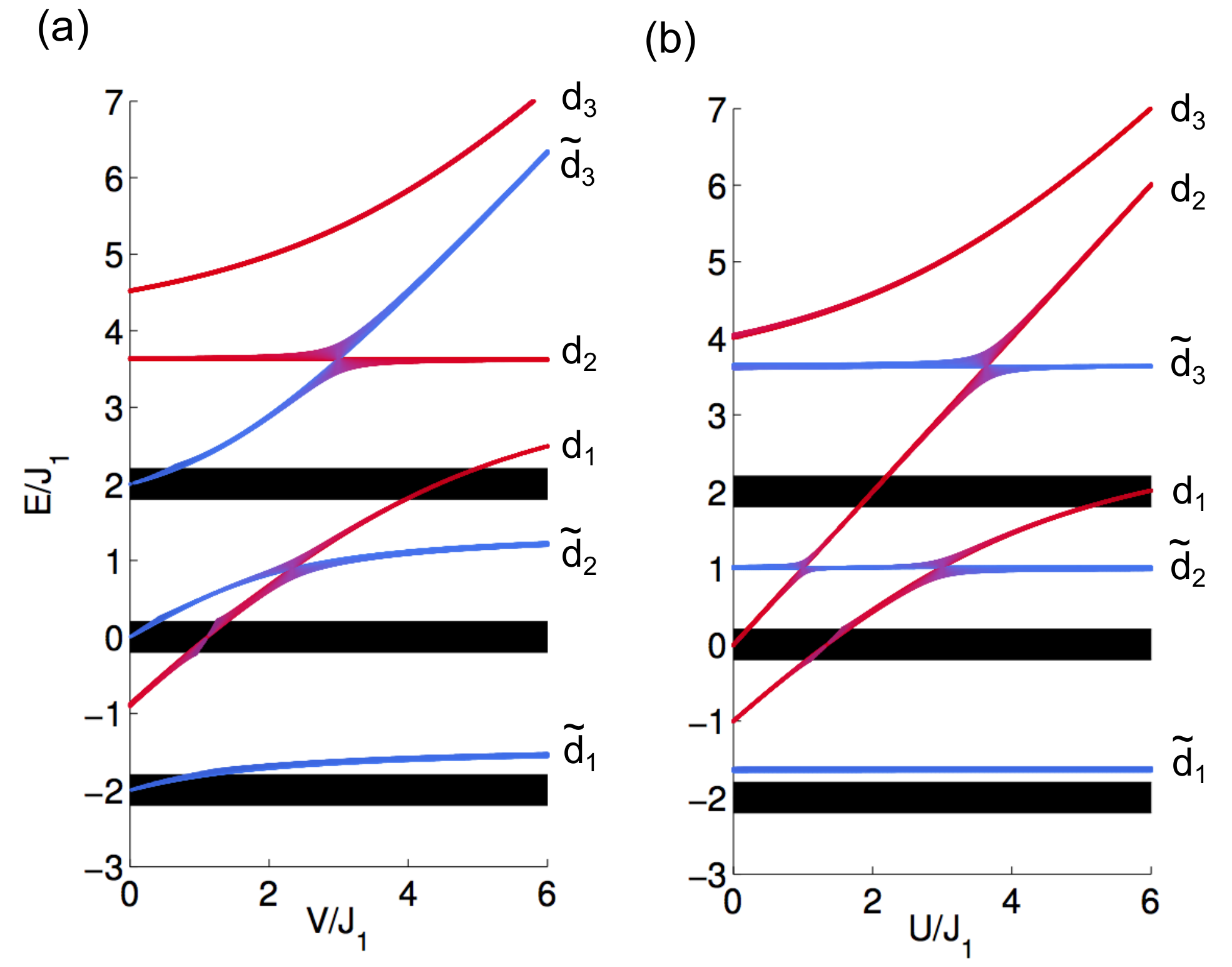}
\caption{(a) Spectrum for $J_2=0.1 J_1$, $U=3.63 J_1$ as a function of $V$
  and (b) spectrum for $J_2=0.1 J_1$, $V=3 J_1$ as a function of $U$
  obtained by exact diagonalization of a lattice of $L=48$ sites
  (24 cells) with PBC. Red colorscale indicates relative in-cell
  population, while blue colorscale indicates relative
  nearest-neighbor out-of-cell population.}
\label{fig:1}     
\end{figure}

\begin{figure}
\center
\includegraphics[width=.65\columnwidth]{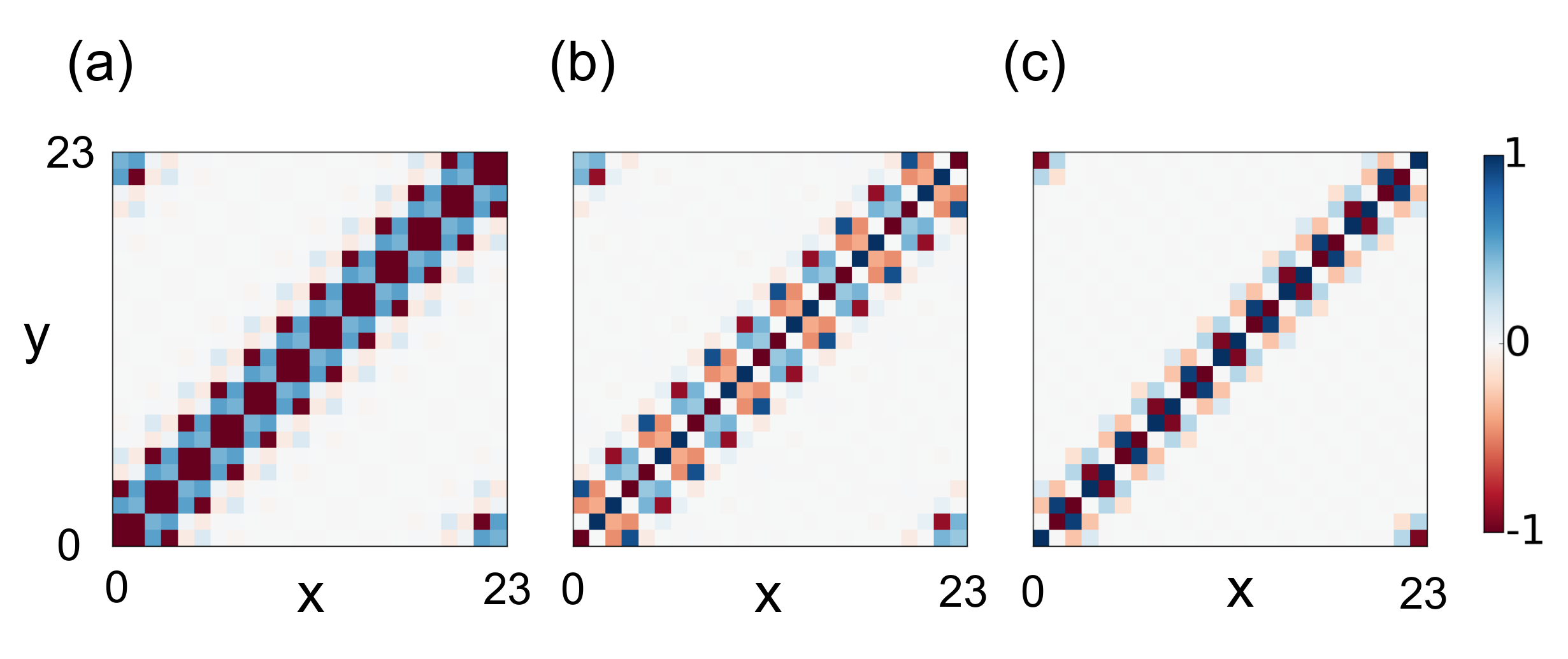}
\caption{Examples of hybridized bound-state wave functions for $V=3
  J_1$ (see Fig.~\ref{fig:1}(b)) at the resonance points (a) $U=J_1$,
  (b) $U=3J_1$ and (c) $U=3.63J_1$.  The color code is normalized to
  the maximum absolute value of the wave function.}
\label{fig:bs-res}      
\end{figure}

To provide some more understanding of the general behaviour of the
system, we plot the spectrum at fixed $U$ and varying $V$ in
Fig.~\ref{fig:1}(a) and the spectrum at fixed $V$ and varying $U$ in
Fig.~\ref{fig:1}(b). The behaviour of the bare dimer energies
$\epsilon_\alpha$ and $\tilde\epsilon_\beta$ can be easily recognized
in those figures.  In Fig.~\ref{fig:1}, one can observe several
crossings occurring. At those points, one of the out-of-cell bound
states $|\tilde d_\beta\ket$ becomes resonant with one of the in-cell
bound states $|d_\alpha\ket$. Similarly to what happens in the uniform
Bose-Hubbard model in the presence of on-site and nearest-neighbors
interactions, at the resonance conditions doublons acquire a much
larger bandwidth. In fact, in contrast to the standard case where
doublons move via second order processes at order $J_2^2$, in-cell
bound states and out-of-cell bound states are now resonantly coupled
at first order in $J_2$. Each single-particle hopping induces the
motion of the center-of-mass of the doublon of half a lattice
cell. This effective reduction of the lattice spacing is translated
into an effective doubling of the Brillouin zone, as observed in
Fig.~\ref{fig:3}(c), where the two bound-state bands relative to
$|d_\alpha\ket$ and $|\tilde d_\beta\ket$ merge into a single one (in
the specific case $\alpha=2$ and $\beta=3$). Correspondingly, the wave
functions are completely hybridized, showing similar in-cell and
out-of-cell populations (see Fig.~\ref{fig:bs-res}(a-c)).

Bound-state energy crossings can be described by an effective
theory. For simplicity of notation, we call $|d\ket$ and $|\tilde
d\ket$ the two in-cell and out-of-cell bound states involved in the
crossing.  Following the standard approach
  \cite{Cohen}, as outlined in the appendix
of \cite{PRA}, we define an effective Hamiltonian $H_{\textrm{eff}}$
that includes $H_{J_2}$ as a perturbation. Hopping between states $|d\ket$
  and $|\tilde d\ket$ takes place at first order in $J_2$ and is
accounted for by the effective hopping matrix elements
\be
J_{\textrm{eff}}\equiv\mv{\tilde d_{i+1/2} |H_{\textrm{eff}}|d_i}
\approx \mv{\tilde d_{i+1/2} |H_{J_2}|d_i} \,.
\label{Jeff}
\ee
Analogously, one obtains $\mv{\tilde d_{i-1/2} |H_{J_2}|d_i} = -
J_{\textrm{eff}}$.  We recall that the subscript $i$ ($i+1/2$)
indicates the center-of-mass of the in-cell (out-of-cell) dimers. For
periodic boundary conditions (PBC), the effective Hamiltonian
parameters do not depend on $i$.  The effective model thus reads
\be H_{\textrm{eff}} = \sum_i \left( \epsilon\, d^\dag_i d^{}_i +
\tilde\epsilon\, \tilde d^\dag_{i+1/2} \tilde d^{}_{i+1/2} \right) +
J_{\textrm{eff}}\sum_i \left( d^\dag_i\tilde d_{i+1/2} - d^\dag_i\tilde
d_{i-1/2} + \textrm{H.c.}\right)\,.
\label{Heff}
\ee
Such effective model describes a single-particle in a superlattice
with alternating potential energy offsets $\epsilon$ and
$\tilde\epsilon$. When the resonance condition
$\epsilon=\tilde\epsilon$ is met, a uniform lattice is obtained,
recovering the case of Fig.~\ref{fig:3}(c). The
  alternating sign of the hopping terms is irrelevant and can be
  reabsorbed through a gauge transformation.

For the specific case of $|d\ket=|d_2\ket$ and $|\tilde d\ket= |\tilde
d_3\ket$, one can use the dimer wavefunctions
\begin{align}
|d_{2,i}\ket &= \f{1}{\sqrt{2}}\left(|A_iA_i\ket - |B_iB_i\ket
\right)\,,\\
|\tilde d_{3,i+1/2}\ket &= \frac{D_1}{\sqrt{2}} \left(|A_iA_{i+1}\ket
+ |B_iB_{i+1}\ket \right) + D_2 |A_iB_{i+1}\ket + D_3 |B_iA_{i+1}\ket \,,
\end{align}
where the coefficients $D_n$ depend on $U$ and $V$ and are obtained
from the diagonalization of Eq.~(\ref{H_nncells}). Hence, the
Hamiltonian parameters read

\begin{eqnarray*}
\epsilon = \epsilon_2\,, \;\;
\tilde\epsilon = \tilde\epsilon_3 \,, \;\;
J_{\textrm{eff}} =  D_3 J_2 \,.
\end{eqnarray*}
This effective theory reproduces exactly the bound state crossing
occurring for $V=3J_1$ around $U \sim 3.63 J_1$.

%%%%%%%%%%%%%%%%%%%%
\section{Edge bound states}
\label{sec:edge}

Let us now consider the case of open boundary conditions (OBC) and
focus our attention on the presence of edge bound states (EBS),
defined as a bound pair localized at the edges of the chain, hence
displaying localization both in the relative and center of mass
coordinates.

\begin{figure}
\center
\includegraphics[width=.75\columnwidth]{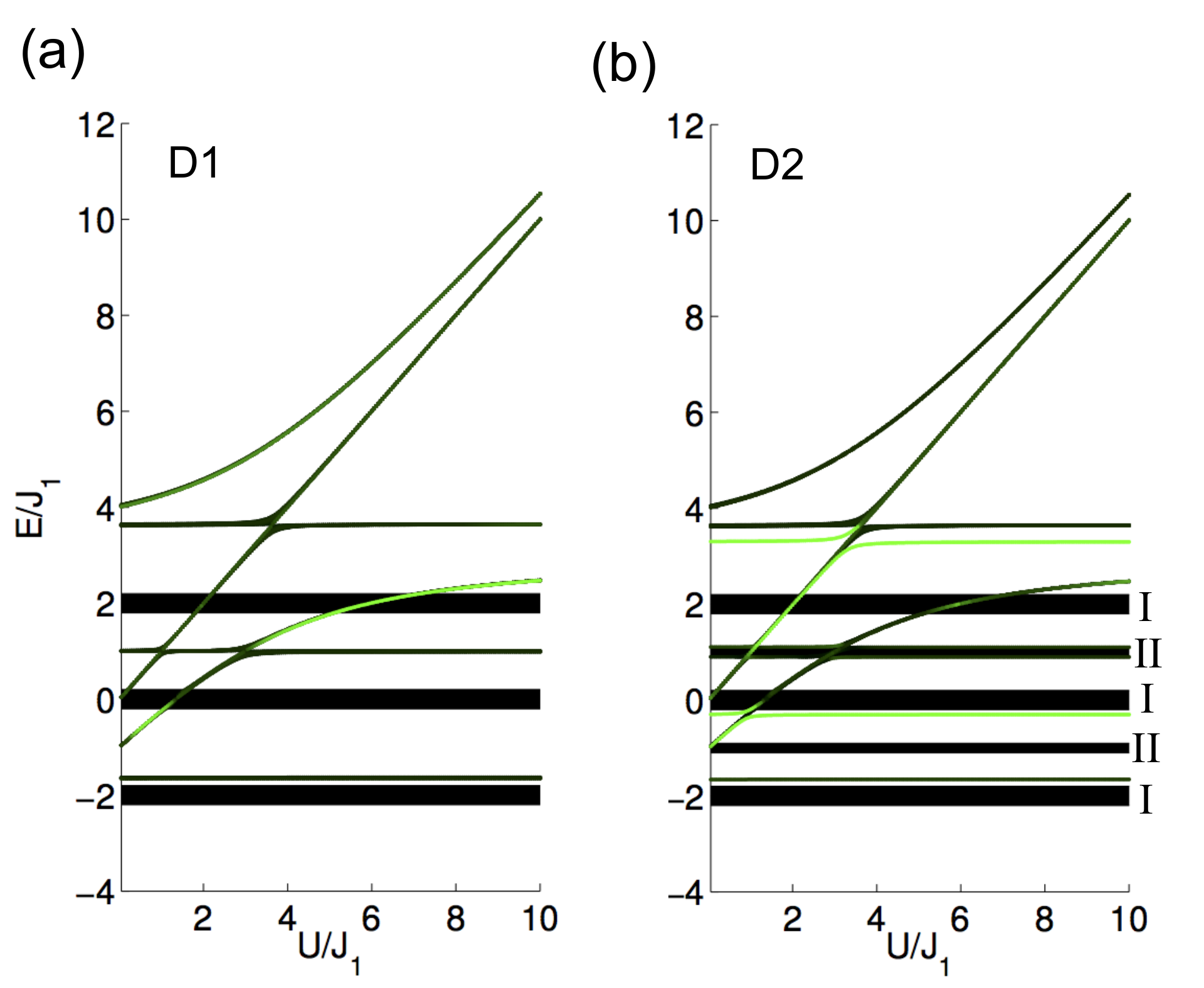}
\caption{Spectrum as a function of $U$ for $J_2=0.1 J_1$ and
  nearest-neighbor interaction $V=3 J_1$ in (a) dimerization $D1$ and
  (b) dimerization $D2$, obtained by exact diagonalization of a
  lattice with OBC for $L=48$ sites (24 cells). Green colorscale
  indicates the relative population in the first 4 lattices sites (2
  lattice cells) which highlights the edge localization of the
  states.}
\label{fig:4}     
\end{figure}

In dimerizations $D1$ (see Fig.~\ref{fig:0}(D1)), the
  existence of the $D1/d_3$ EBS for small $U$ is confirmed also when
  nearest-neighbor interactions are present.  The most noteworthy
  feature introduced by $V$ is a novel strongly localized EBS related
  to bound state $d_1$ (see lower green line in
  Fig.~\ref{fig:4}(a)). This state, absent for $V=0$, exists for
  generic $U \gtrsim V >0$. 
Notice that there is an interesting correspondence between
  $D1/d_3$ EBS at small $U$ and $D1/d_1$ EBS at large $U$. In fact, in
  both cases the wavefunction is related to dimer state $|A_iB_i\ket$,
  which possesses large energy $V$ for $0< U \ll V$ and small energy
  $V$ for $J_1 \ll V \ll U$. For increasing $U$ or $V$ respectively,
  admixture with dimer states $|A_iA_i\ket$ and $|B_iB_i\ket$ occurs
  and the EBS eventually lose localization.
The origin of $D1/d_1$ and $D1/d_3$ EBS will be explained based on a
Tamm-like model at the end of this section.

In dimerization $D2$ (see Fig.~\ref{fig:0}(D2)), as clearly visible by
the green lines in Fig.~\ref{fig:4}(b), very pronounced EBS appear.
At small $V$, these EBS arise at energy close to $\pm J_1$ and are
reminiscent of the $D2/d_{1,2}$ EBS discussed in \cite{PRA}, namely
on-site edge doublons undergoing a mixing with a single-particle edge
state plus a free scattering particle (denoted as type II scattering
continuum).
These EBS can be predicted by the same reduced theory developed in
\cite{PRA} (see matrix $\cal H^{\textrm{red}}$ in Eq. (A25)), upon
introducing the appropriate corrections due to nearest-neighbor
interactions. Apart from on-site edge doublons and type II scattering
states, the basis of the reduced theory has now to contain out-of-cell
edge doublons. Hence, the starting point of the reduced theory is a
proper truncation ansatz for the states at the edges.  For instance,
in the specific case of states $|d_2\ket$ and $|\tilde d_3\ket$, at
the left edge of the lattice one can define
\be |d_{2,0}\ket = - |B_0 B_0\ket \,, \quad |\tilde d_{3,1/2}\ket =
D^{L}_3 |B_0A_{1}\ket + D^{L}_1 |B_0B_{1}\ket \,.
\label{ansatz_edge}
\ee
While state $|d_{2,0}\ket$ is the same one considered in \cite{PRA},
state $|\tilde d_{3,1/2}\ket$ differs from $|B_0\ket \otimes
(|A_{1}\ket -|B_{1}\ket)$ by a renormalization of the coefficients due
to $V$.  More precisely, $D^{L}_1$ and $D^{L}_3$ depend on $V$ and are
obtained from the eigenvector of the matrix
\be
\label{hamNN}
H^{\textrm{nn-cells}}_{1/2} =
\begin{pmatrix}
V & - J_1\\
-J_1 & 0
\end{pmatrix}\,
\ee
corresponding to the largest eigenvalue $\lambda_+ =
(V+\sqrt{V^2+4J_1^2})/2$.  Consequently, the matrix $\mathcal
H^{\textrm{red}}$ is identical to the one in Eq.~(A25) of
Ref.~\cite{PRA}, apart for the elements involving $|\tilde
d_{3,1/2}\ket$. The upper left corner of $\mathcal H^{\textrm{red}}$,
which describes the left edge physics, becomes
\be
\mathcal H^{\textrm{red}}_L=
\left( 
\begin{array}{ccc}
 U &  -\sqrt 2 J_2 D^L_3 &  0 \\ 
 -\sqrt 2 J_2 D^L_3& \lambda_+ & - J_2 D^L_1/\sqrt{2}     \\
 0 &- J_2 D^L_1/\sqrt{2}& J_1 
\end{array}
\right)\,.
\label{megamatrix}
\ee
An analogous treatment holds for the right edge, modifying the lower
right corner of $\mathcal H^{\textrm{red}}$. This modified theory
predicts the existence of two possible kinds of EBS, namely $D2/d_2$
and the novel $D2/\tilde d_3$ (upper green diagonal and horizontal
lines in Fig.~\ref{fig:4}(b), respectively). The new feature, due to
nearest-neighbor interactions, is the presence of the $D2/\tilde d_3$
EBS.
For $V=0$, the $D2/d_2$ EBS smoothly transforms into a type II
scattering state as $U$ moves away from $J_1$ (see Ref.~\cite{PRA},
Fig.~7). Instead when scanning $U$ in the presence of $V\neq 0$, the
$D2/d_2$ EBS smoothly becomes a $D2/\tilde d_3$ EBS. Near the avoided
crossing, the two EBS hybridize.  The $[2(L-1)+3]\times[2(L-1)+3]$
matrix $\mathcal H^{\textrm{red}}$ (where $L$ is the number of cells
in the lattice) correctly accounts for the energy shift of the
$D2/\tilde d_3$ EBS above type II scattering continuum even for very
small values of $V$.  The effect of $V$ is to bring the two-body edge
components of the edge/free particle states out of resonance with the
rest of type II scattering continuum. Eventually, for sufficiently
large $V$, $D^L_1\ra 0$, such that $|d_{2,0}\ket$ and $|\tilde
d_{3,0}\ket$ decouple from the type II continuum. One can therefore
simplify the description provided by the complete $\mathcal
H^{\textrm{red}}$ matrix by taking only the first $2 \times 2$ block
of $\mathcal H^{\textrm{red}}_L$ in Eq. (\ref{megamatrix}) for the
left edge, and an analogous $2 \times 2$ block for the right edge.

\begin{figure}
\center
\includegraphics[width=.7\columnwidth]{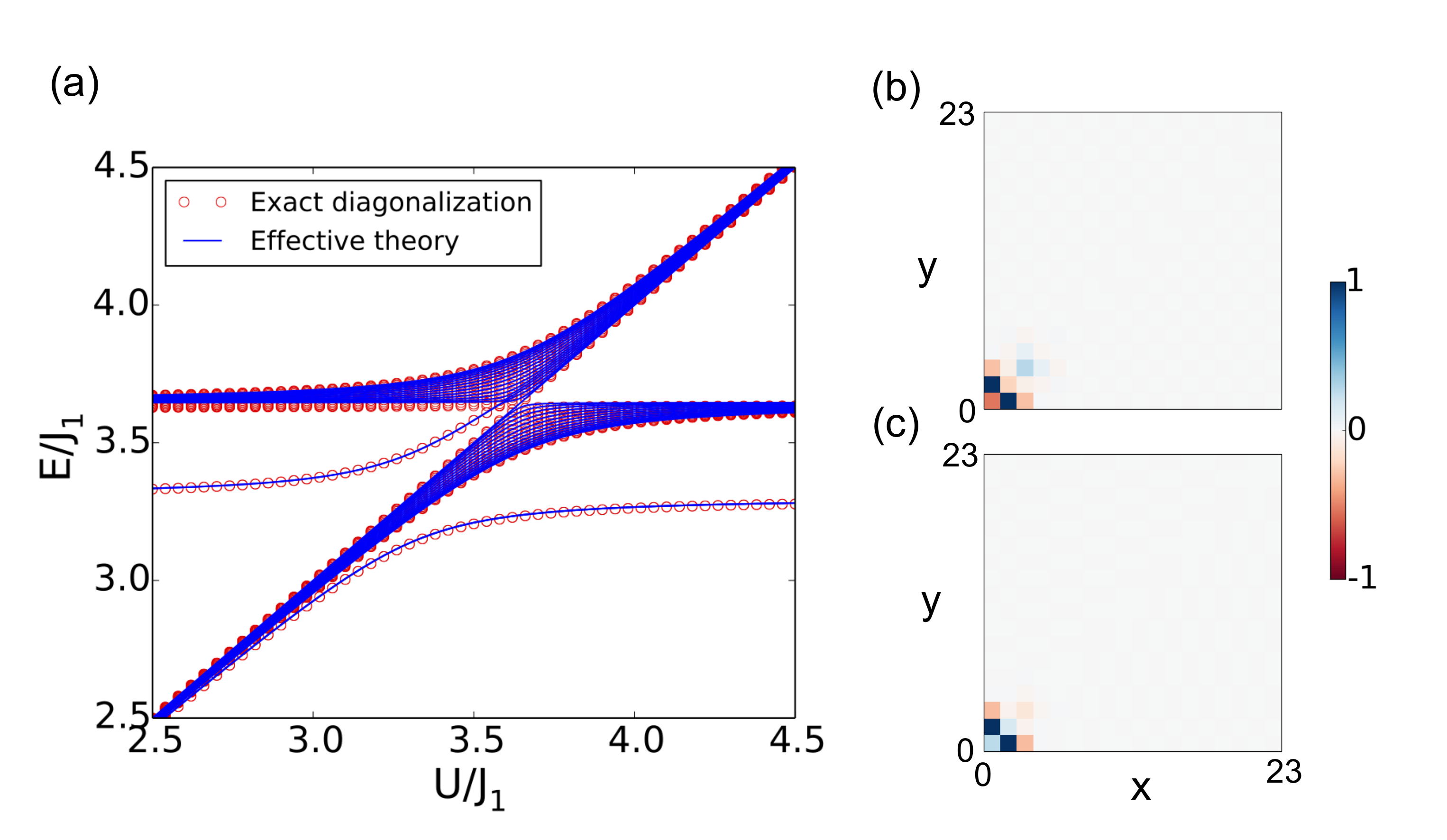}
\caption{(a) Closer view of the spectrum with OBC in dimerization $D2$
  with the same parameters as in Fig.~\ref{fig:4}(b) for $36$ lattice sites. 
  A comparison between the exact spectrum and
  the one obtained from the effective theory in
  (\ref{Heff}),(\ref{ren_par}) is shown. (b-c) Edge bound states for
  $U=3J_1$ (b) and $U=4J_1$ (c). The color code is normalized to the
  maximum absolute value of the wave function.}
\label{fig:5}     
\end{figure}

The accuracy of the $2\times 2$ model proves excellent in reproducing
the avoided crossing in Fig.~\ref{fig:4}(b) (upper green curves). For
$ V\sim U$, namely when nearest-neighbor interactions and on-site
interactions are of the same order, the presence of the $d_2$ and
$\tilde d_3$ bound state narrow bands becomes relevant. The $D2/d_2$
EBS predicted with the reduced theory has no numerical evidence for
sufficiently large $U$, such that $\epsilon_2 \gtrsim \tilde
\epsilon_3$. To include the effect of the $d_2$ and $\tilde d_3$ bound
state narrow bands, we can resort to the effective model presented in
Eq.~(\ref{Heff}).  Considering the modified states at the edges as in
Eq.~(\ref{ansatz_edge}), one finds
\be
\epsilon^{\textrm{edge}} = \epsilon = U \,,\,\,\,\,
\tilde\epsilon^{\textrm{edge}} = \frac{ V+\sqrt{V^2 + 4J_1^2} }{2}
\neq\tilde\epsilon \,, \,\,\,\, J^{\textrm{edge}}_{\textrm{eff}} =
\sqrt 2 D^{L}_3 J_2\,,
\label{ren_par}
\ee
providing a substantial renormalization of the edge parameters with
respect to the bulk\footnote{Since the basis states
  (\ref{ansatz_edge}) are in common between the effective models
  (\ref{megamatrix}) and (\ref{Heff}),(\ref{ren_par}), it is not a
  coincidence that the renormalized parameters in (\ref{ren_par})
  coincide with the matrix elements of (\ref{megamatrix}).}. In this
respect, the emerging EBS can be interpreted as Tamm-like states. When
$U$ is sufficiently far from the value where the crossing takes place
and hybridization with other states does not occur, the $D2/\tilde
d_3$ EBS arises mainly thanks to $\tilde \epsilon^{\textrm{edge}}$ and
its energy becomes independent of $U$.
As shown in Fig.~\ref{fig:5}(a), around the level crossing the
agreement between effective model and numerical results is
excellent. Since higher type II scattering continuum and bound state
$d_3$ are not included in model (\ref{Heff}),(\ref{ren_par}), the
conditions required for an accurate prediction of the bound state
bandwidth and the EBS energies are $U \sim V$ and $V>2J_2$,
respectively.  At fixed $U$ and for large $V$, the $D2/{\tilde d}_3$
EBS persists, preserving excellent localization properties. This is
due to the fact that, even if its energy gap with respect to the
$|{\tilde d}_3\ket$ bound-state narrow band becomes smaller and
smaller, the ratio between the gap and the $|{\tilde d}_3\ket$
bandwidth remains finite, preventing EBS diffusion.

\begin{figure}
\center
\includegraphics[width=.95\columnwidth]{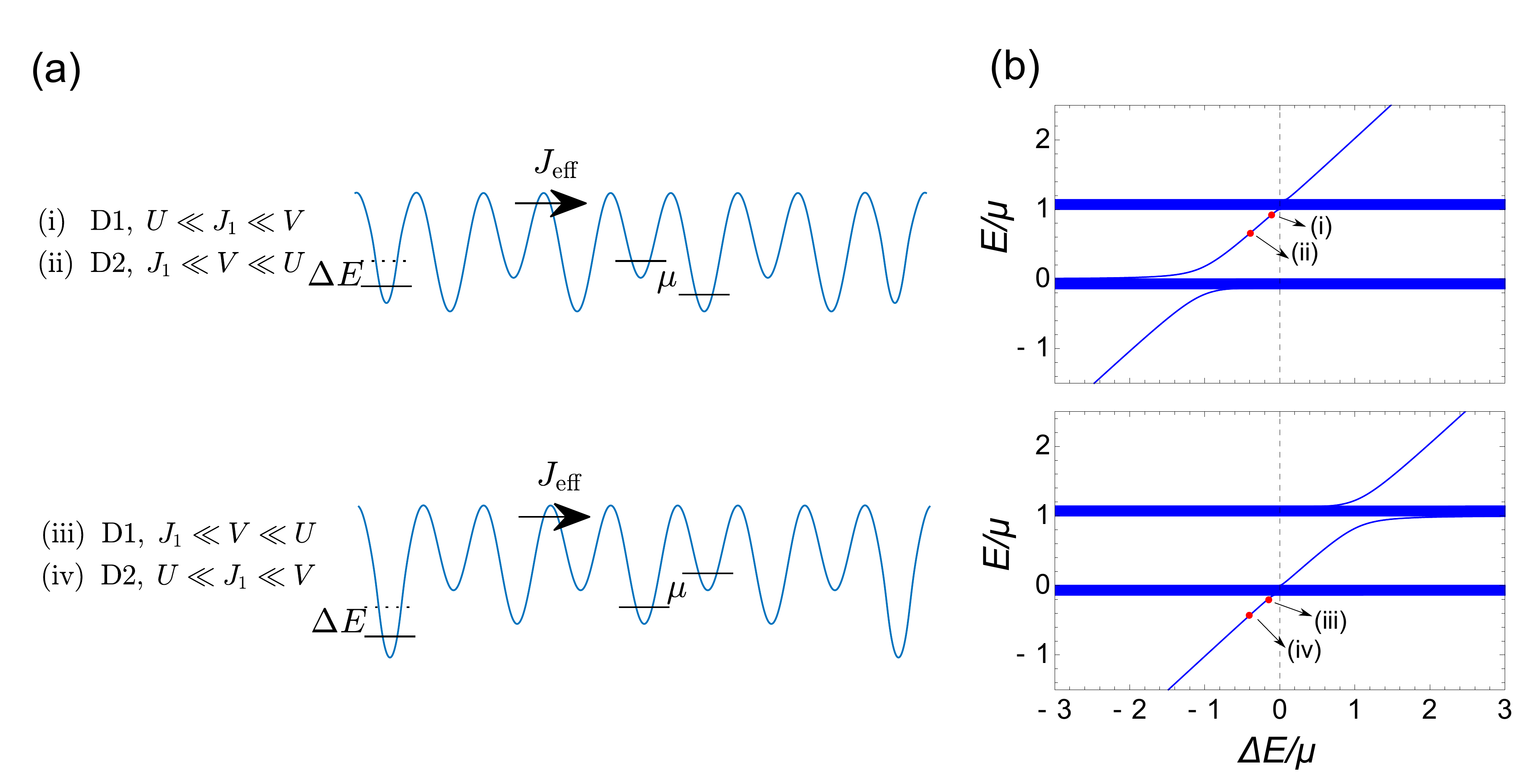}
\caption{(a) Effective single-particle ionic Hubbard model with
  hopping $J_{\rm eff}$, potential energy off-set $\mu$, and edge
  potential energy correction $\Delta E$; The boundary conditions in
  the upper panels accounts for the cases (i)-(ii), while the boundary
  conditions in the lower panels account for the cases (iii)-(iv), as
  indicated.  (b) Tamm analysis for the effective ionic Hubbard model
  for fixed $\mu$ and $J_{\rm eff}$ and varying $\Delta E$; The zero
  of the energy is set to the lowest bulk energy offset; The red dots
  indicate the prediction for the edge modes for the four cases: (i)
  $D1/d_3$ EBS just below the upper ($d_3$) continuum for $U \ll J_1
  \ll V$; (ii) novel strongly localized $D2/\tilde d_3$ EBS in between the lower ($d_1$)
  and the upper ($\tilde d_3$) continua with gap $\mu/2$ for $J_1 \ll
  V \ll U$; (iii) novel $D1/d_1$ EBS just below the lower ($d_1$)
  continuum for $J_1 \ll V \ll U$; (iv) novel strongly localized
  $D2/\tilde d_3$ EBS below the lower ($\tilde d_3$) continuum with
  gap $\mu/2$ for $U \ll J_1 \ll V$.}
 \label{fig:7}      
\end{figure}

A unified picture to describe the novel $D1/d_1$ and $D2/\tilde d_3$
EBS can be obtained resorting to an
  effective model in the strongly interacting regime. For both $D1$
and $D2$, the states involved are $d_3$ and $\tilde d_3$ in the limit
$U \ll J_1 \ll V$ and $d_1$ and $\tilde d_3$ in the limit $J_1 \ll V
\ll U$.
Dimers states $|d_3\ket$ for $U \ll V$ and $|d_1\ket$ for $U \gg V$
are accounted for by the same ansatz $|A_iB_i\ket$, and dimer $|\tilde
d_3\ket$ is approximated by $|A_iB_{i+1}\ket$.
Contrary to the effective model presented in Sec.~\ref{sec:bound},
both hopping terms $J_1$ and $J_2$ are now considered at the
pertubative level, through second order processes.  One obtains an
effective single-particle ionic Hubbard model with constant hopping
parameter $J_{\rm eff}=J_1J_2 (1/V+2/(V-U))$ and alternating potential
energy offsets.
The underlying effective lattice sites correspond to pairs of sites of
the original lattice. The energy offsets in the effective lattice are
given by the energy of the corresponding (either in-cell or
out-of-cell) dimer states in the original lattice, leading to a
site-to-site energy difference $\mu = 2 (J_1^2-J_2^2)|1/V-2/(V-U)|$.
Depending on whether the energy offset at the boundary of the
effective lattice is the largest or the
smallest (which depends at once on the dimerization $D1$ or $D2$ of
the original lattice and on the relation between $U$ and $V$), the
four cases under consideration are represented by one or the other
effective ionic Hubbard model sketched in Fig.~\ref{fig:7}(a) (see
labels in figure for a better understanding).
Moreover, depending on the dimerization $D1$ or $D2$ of the original
lattice, open boundary conditions introduce at the outmost
sites a further energy off-set
$\Delta E$, as sketched in Fig.~\ref{fig:7}(a).

The spectrum of the ionic Hubbard model presents two continua
respectively below and above energies $0$ and $\mu$, with bandwidth
${\cal W}=|\mu-\sqrt{\mu^2+16J_{\rm eff}^2}|/2$.  Depending on $\Delta
E$ and on the boundary conditions, the Tamm-analysis generalized to
the ionic Hubbard model, shown in Fig.~\ref{fig:7}(b), predicts that
the system can support gapped edge modes either below or above the
continua, or in the gap (see Fig.~\ref{fig:7}(b)).

In dimerization $D1$, both for $U \ll J_1 \ll V$ and $J_1 \ll V \ll
U$, one obtains $\Delta E = -J_2^2/V$, which is much smaller than
$\mu$ in the limit $J_2 \ll J_1$.
This energy correction, even if very small, allows for edge modes just
below the upper continuum (i) and even below the lower continuum
(iii). In case (i), the Tamm analysis reveals that any small $\Delta E
<0$ is sufficient to localize states at the edges. In case (iii), the
required condition is $|\Delta E| > {\cal W}/2$. A perturbative
calculation shows that the negative edge-energy off-set $\Delta E$ is
sufficiently large in absolute value to overcome the bandwidth of the
lower continuum. Indeed, at second order in $U/V$ and $J_2/J_1$, one
finds ${\cal W}=2|\Delta E| [1-6V/U+10 (V/U)^2]$, such that edge modes
are present for all values of $U$. However, while for $V/U \gtrsim
0.1$ the edge modes are strongly localized, for larger $U$ one
observes a crossover to states with very large localization length
because $|\Delta E|\ra {\cal W}/2$.

In dimerization $D2$, both for $U \ll J_1 \ll V$ and $J_1 \ll V \ll
U$, one obtains $\Delta E = -J_1^2/V$, such that in the limit $J_2 \ll
J_1$, $\Delta E \approx - \mu/2$. This substantial edge correction is
at the origin of the strongly localized edge modes that appear in
between the two
continua (ii) or well below the lower continuum (iv).

These predictions exactly agree with the numerical findings in the
appropriate interaction regimes. Moreover, the effective
single-particle ionic Hubbard Tamm analysis distinctly highlights the
role of nearest-neighbor interactions behind the existence and
the localization properties of the $D1/d_3$, $D1/d_1$,
$D2/\tilde d_3$ EBS.

While it might seem intuitive that strong interactions localize, we
recall that this is in general not true. In the case of on-site
interactions only, we have proven that EBS localization is not
supported in a finite system in the limit $U \to \infty$ \cite{PRA}.
Here, we demonstrate that the presence of nearest-neighbor
interactions guarantees the existence of strongly-localized EBS in
both dimerizations even for large values of $U$ and $V$.

%%%%%%%%%%%%%%%%%%%%%%%
\section{Conclusions}
\label{sec:conclusions}

In this work, we have discussed the two-body problem in a
Su-Schrieffer-Heeger chain with on-site and nearest-neighbor
interactions, extending the results presented in Ref.~\cite{PRA}. We
have identified optical fiber setups as the most promising
experimental realization of our predictions, thanks to the possibility
of implementing large nearest-neighbor interactions. We have shown
that two types of bound states exist: in-cell bound states analogous
to the ones found in Ref.~\cite{PRA} in the presence of on-site
interactions only, and out-of-cell bound states that exist thanks to
nearest-neighbor interactions. These two types of dimers become
resonant depending on the values of $U$ and $V$, yielding bound states
with mixed in-cell/out-of-cell character and larger mobility.

We have then considered open boundary conditions.  We have numerically
found that nearest-neighbor interactions $V$ generate
strongly-localized $D1/d_1$ EBS, which do not exist in dimerization
$D1$ at $V = 0$.
In dimerization $D2$, we have studied the fate of the
two-body edge-bound state $D2/d_2$ in the presence of the novel $D2/\tilde d_3$
EBS, created by nearest-neighbor interactions.
When the energies of $D2/d_2$ and $D2/{\tilde d}_3$ EBS are
comparable, the two edge states become hybridized. We have provided a
careful characterization of the energy spectrum at the crossing.

In the strongly-interacting limit, both $D1/d_1$ and $D2/\tilde d_3$
EBS can be interpreted as Tamm-like states of an effective
single-particle ionic Hubbard model.  This model explains why both the
$D1/d_1$ EBS and the $D2/\tilde d_3$ EBS persist even at very large
values of on-site and nearest-neighbor interactions, preserving
excellent localization properties.  The presence of strongly-localized
EBS for large values of interactions is a highly non-trivial result of
our work.

%%%%%%%%%%%%%%%%%%%%%%%%%%
\section{Acknowledgements}

A.R. acknowledges support from the Alexander von Humboldt foundation
and W.~Zwerger for the kind hospitality at the TUM. This work was
supportedby the EU-FET Proactive grant AQuS, Project No.  640800 and
by Provincia Autonoma di Trento, partly through the project SiQuro.

\end{document}